\documentclass[letterpaper,10pt,conference]{IEEEtran}
\IEEEoverridecommandlockouts

\usepackage{cite}
\usepackage{amsmath,amssymb,amsfonts}
\usepackage{graphicx}
\usepackage{textcomp}
\usepackage{xcolor}
\usepackage{booktabs}
\usepackage{multirow}
\usepackage[hidelinks]{hyperref}
\usepackage[nameinlink,capitalize]{cleveref}
\usepackage{orcidlink}
\def\BibTeX{{\rm B\kern-.05em{\sc i\kern-.025em b}\kern-.08em
    T\kern-.1667em\lower.7ex\hbox{E}\kern-.125emX}}

\begin{document}

\title{Parameter-Efficient Quantum-Inspired Fast Weight Programmers for Traffic-Matrix Forecasting\thanks{The views expressed in this article are those of the authors and do not represent the views of Wells Fargo. This article is for informational purposes only. Nothing contained in this article should be construed as investment advice. Wells Fargo makes no express or implied warranties and expressly disclaims all legal, tax, and accounting implications related to this article.}}

\author{
\IEEEauthorblockN{
    Kuo-Chung Peng$^{1,2,}$\orcidlink{0009-0001-8342-2481},
     Jiun-Cheng Jiang$^{1,2}$\IEEEauthorrefmark{1}\orcidlink{0009-0005-1134-4962},
     Chun-Hua Lin$^{1,2}$\orcidlink{0009-0002-4383-0453}, \\
     Tai-Yue Li$^{2}$\orcidlink{0000-0002-1993-1863},
     Nan-Yow Chen$^{2,}$\IEEEauthorrefmark{2}\orcidlink{0000-0001-8139-6809},
    Samuel Yen-Chi Chen$^{3,}$\IEEEauthorrefmark{3}\orcidlink{0000-0003-0114-4826}
}
\IEEEauthorblockA{$^1$Department of Physics and Center for Theoretical Physics, National Taiwan University, Taipei, Taiwan}
\IEEEauthorblockA{$^2$National Center for High-Performance Computing, National Institutes of Applied Research, Hsinchu, Taiwan}
\IEEEauthorblockA{$^3$Wells Fargo, New York, NY, USA}

\IEEEauthorblockA{Emails: 
\IEEEauthorrefmark{1}\href{mailto:jcjiang@phys.ntu.edu.tw}{jcjiang@phys.ntu.edu.tw}, 
\IEEEauthorrefmark{2}\href{mailto:nanyow@nchc.narl.org.tw}{nanyow@nchc.narl.org.tw},
\IEEEauthorrefmark{3}\href{mailto:ycchen1989@ieee.org}{ycchen1989@ieee.org}
}
}
\maketitle

\begin{abstract}
Traffic matrices (TMs) capture network-wide origin--destination demand and are central to traffic engineering, yet accurate whole-matrix forecasting remains challenging when prediction must be performed under the memory, update, and training-budget constraints of online network control. This paper investigates whether compact quantum-inspired recurrent models can provide effective TM forecasts without relying on dedicated graph, transformer, or diffusion modules. We adapt gated quantum-inspired Kolmogorov--Arnold network fast-weight programmers (QKAN-FWPs) to direct multi-step Abilene TM forecasting, where each model predicts the next 20 five-minute frames of a 144-channel origin--destination (OD) matrix from a two-hour history. We benchmark three QKAN placement variants against a matched-size long short-term memory (LSTM) network, a larger LSTM, and a classical gated fast-weight programmer under a shared fixed-budget training protocol. Among the evaluated recurrent models, G-QKANFWP achieves the best pooled root-mean-square error (RMSE), while using only $22.4\%$ of the larger LSTM. It also outperforms both the matched-size LSTM and the classical G-FWP baseline, indicating that the gain is not due to gated fast-weight framework alone. Convergence and channel-wise analyses further show that the quantum-inspired variants obtain lower validation-loss area under the learning curve (AULC) than matched-size recurrent baselines, while G-QKANFWP and GQKAN-FWP achieve substantially more OD-channel wins. These results identify a classical slow programmer with a quantum-inspired fast programmer as a promising accuracy--efficiency design for resource-conscious network traffic-matrix forecasting.
\end{abstract}


\section{Introduction}

Traffic matrices (TMs) summarize traffic demand between ingress--egress, or origin--destination (OD), node pairs and are a standard representation for network measurement, tomography, and traffic engineering~\cite{vardi1996network,zhang2003fast,medina2002traffic,tune2013internet}. In backbone traces such as Abilene and G\'EANT~\cite{Abilene_data}, each TM is naturally a high-dimensional observation: OD entries evolve over time while also reflecting network-wide demand patterns. Forecasting future TMs is therefore more structured than predicting an isolated link or scalar flow. A useful forecaster must model temporal dynamics while preserving cross-channel dependencies that are important for capacity planning, congestion mitigation, routing, failure response, and resource scheduling.

Recent traffic-forecasting methods increasingly exploit spatio-temporal structure through neural sequence models, graph or diffusion operators, image-like TM encodings, and generative formulations~\cite{liu2019traffic,azzouni2018neutm,kablaoui2024trafficimage,sun2026lead}. These approaches have improved predictive accuracy, but many are designed under regimes where model capacity and runtime are not the primary limitation. In operational networks, forecasting is often embedded in online control loops, where models may need to be trained, updated, or executed within short routing or orchestration intervals. This is especially relevant in edge and cloud-edge settings: moving fine-grained telemetry to a remote cloud can add bandwidth cost and decision latency, while local prediction places pressure on memory, compute, and energy budgets~\cite{huo2021cloudedge,ferreira2023survey,perifanis2023energyaware,hu2024edgecloud,zhu2025lightweight}. For such settings, parameter count and fixed-budget convergence are part of the modeling objective rather than implementation details.

This paper asks whether compact quantum-inspired recurrent models can provide effective whole-matrix TM forecasts under this resource-conscious view. We adapt the gated quantum-inspired Kolmogorov--Arnold network fast-weight programmer (QKAN-FWP) family~\cite{peng2026gatedqkanfwp} to direct multi-step Abilene TM forecasting. Prior gated QKAN-FWP work studied parameter-efficient sequence modeling with dynamically updated fast weights on single-step time series prediction, direct multi step sunspot number forecasting, and reinforcement learning tasks; here, we test whether the same mechanism remains useful when the output is a 144-channel traffic matrix with coupled OD structure. We deliberately focus on compact recurrent models rather than adding an explicit graph, vision, transformer, or diffusion module. This choice isolates the effect of the gated quantum-inspired fast-weight design and positions the study as an accuracy--efficiency evaluation rather than a claim to replace high-capacity TM forecasters.

Our contributions are as follows:
\begin{enumerate}
\item We benchmark three gated QKAN-FWP variants against a matched-size long short-term memory (LSTM) network, a larger LSTM, and a classical G-FWP baseline on a multi-channel TM forecasting task.
\item We show that G-QKANFWP achieves the best pooled root-mean-square error (RMSE) among the evaluated recurrent models, outperforms both LSTM-S and the classical G-FWP baseline, and slightly improves over LSTM-L while using only 22.4\% of LSTM-L's parameters.
\item We complement aggregate accuracy with convergence and OD-channel analyses: all quantum-inspired variants achieve validation-loss area under the learning curve (Val-loss AULC) under the fixed epoch budget than LSTM-S and G-FWP, while G-QKANFWP and GQKAN-FWP win substantially more OD channels than these matched recurrent baselines.
\end{enumerate}

\section{Related Work}

\textbf{Traffic-matrix and spatio-temporal forecasting.}
Classical TM research studied how to estimate OD demand from partial network measurements and link loads~\cite{vardi1996network,medina2002traffic,zhang2003fast,tune2013internet}. Forecasting extends this objective from estimating current or historical demand to predicting future network states. Early and neural traffic predictors include deep-learning approaches for dynamic traffic engineering and SDN-based TM prediction~\cite{liu2019traffic,azzouni2018neutm}. More recent work draws on the broader spatio-temporal forecasting literature, where DCRNN models directed traffic flow with diffusion convolution, STGCN uses spatio-temporal graph convolutional blocks, and Graph WaveNet learns adaptive dependencies with dilated temporal convolutions~\cite{li2017dcrnn,yu2017stgcn,wu2019graphwavenet}. TM forecasting differs from road-sensor forecasting because channels correspond to OD flows rather than physical sensors, but both settings require joint temporal and cross-channel modeling. Recent TM methods therefore also explore image-like representations, graph or diffusion modules, and conditional generative models~\cite{kablaoui2024trafficimage,sun2026lead}. Our work is complementary: instead of proposing a new spatial encoder, we evaluate whether compact recurrent fast-weight models can deliver favorable accuracy--parameter trade-offs for whole-matrix prediction.

\textbf{Compact sequence models.}
Recurrent neural networks, including LSTMs~\cite{hochreiter1997lstm}, remain attractive for streaming prediction because they process sequences incrementally. Fast-weight programmers provide an alternative recurrent mechanism by storing temporal information in dynamically updated parameters rather than only in hidden states~\cite{schmidhuber1992fastweights,schlag2021linear,irie2021recurrentfwp}. Quantum and quantum-inspired sequence learning has also been explored through quantum reservoir computing~\cite{mujal2023timeqrc,kobayashi2024feedbackqrc, chen2024efficient}, quantum LSTM variants~\cite{chen2022qlstm,khan2024qlstm,hsu2026qkanlstm,lin2024quantum,hsu2025quantum}, hybrid quantum models~\cite{chen2025qrwkv,choudhary2026hqnn,rivera2022time,liu2025quantum}, and quantum fast weight programmers~\cite{chen2024qfwp,liu2025programming,ceschini2026quantum}. QKANs use quantum variational activation functions as compact nonlinear modules inside Kolmogorov--Arnold networks~\cite{jiang2025qkan}, while gated QKAN-FWP integrates QKAN modules with gated fast-weight programming framework~\cite{peng2026gatedqkanfwp}. This paper extends that family from prior sequence-learning settings to spatially coupled, multi-channel TM forecasting and compares where QKAN placement is most effective within the fast-weight architecture.

\section{Model Architectures and Baselines}

Fast-weight programmers replace a purely hidden-state recurrence with a compact set of dynamically updated fast parameters. A slow pathway reads the current input and proposes an update to these fast parameters; a fast pathway then uses the current fast parameters to generate the sequence output. 
In the gated variants, a scalar gate interpolates between the
previous fast parameters and the new proposal. This gate acts as a lightweight stabilizer for parameter evolution, while the fast parameters carry temporal information through the sequence~\cite{peng2026gatedqkanfwp}. The Gated QKAN-FWP model uses Hybrid QKAN (HQKAN), an encoder--processor--decoder instantiation of Jiang--Huang--Chen--Goan network (JHCG Net)~\cite{jiang2025qkan,jiang2025qkan_github}, where a classical encoder forms latent features, a QKAN block transforms them nonlinearly, and a decoder generates the output~\cite{peng2026gatedqkanfwp}. 

\begin{figure}[t]
\centering
\includegraphics[width=0.95\columnwidth, trim={0cm 1cm 0cm 0cm}, clip]{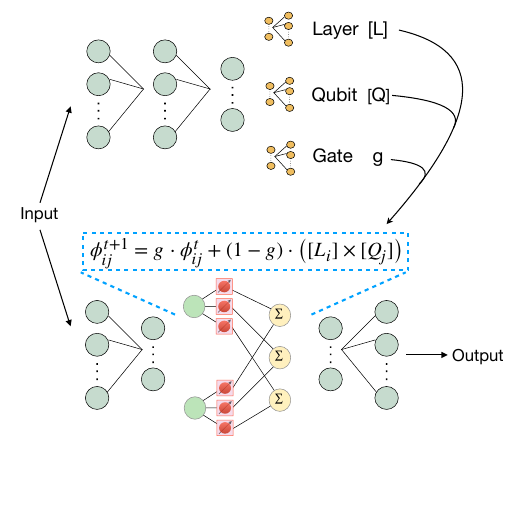}
\vspace{-15pt}
\caption{\textbf{Architecture of G-QKANFWP.} A classical slow programmer dynamically generates the parameters of an HQKAN fast programmer.}
\label{fig:g-qkanfwp}
\vspace{-20pt}
\end{figure}

\textbf{G-QKANFWP} uses a classical slow programmer and places the HQKAN nonlinearity in the fast readout, as illustrated in \cref{fig:g-qkanfwp}. This variant tests whether a quantum-inspired fast programmer is useful when the update generator is kept simple.

\textbf{GQKAN-FWP} places the quantum-inspired module in the slow programmer and uses a classical linear fast programmer. This variant tests whether HQKAN is more useful before the fast memory is generated than inside the fast readout.

\textbf{GQKAN-QKANFWP} uses HQKAN modules on both the slow and fast sides. It is the smallest model in the comparison and tests whether more aggressive compression can preserve whole-matrix forecasting accuracy.

The classical \textbf{G-FWP} baseline keeps the gated fast-weight framework but removes the HQKAN components \cite{peng2026gatedqkanfwp}. This baseline is important because it separates the benefit of gated fast-weight framework from
the benefit of the quantum-inspired modules. We also include two LSTMs: LSTM-S, a matched-size baseline, and LSTM-L as a larger baseline.

\section{Data and Experimental Protocol}

We use the Abilene traffic-matrix (TM) dataset~\cite{Abilene_data}, collected from the Internet2 backbone network. Following the Abilene whole-matrix forecasting protocol in~\cite{sun2026lead}, we process the 24 weekly files as five-minute TM frames. Each row contains 720 values arranged as 144 OD pairs with five values per pair. We select the real OD-traffic value from each group, yielding $24\times2016=48{,}384$ frames, each represented as a 144-dimensional OD vector. The $12\times12$ matrix form is used only for
visualization.

Before training and evaluation, each frame is converted to the frame-normalized TM (FN-TM) representation. For an OD entry $x_{ij}^{(\tau)}$ in frame $\tau$, we compute
\begin{equation}
 v_{ij}^{(\tau)} =
 1-\frac{x_{ij}^{(\tau)}-\min_{k,\ell}x_{k\ell}^{(\tau)}}
 {\max_{k,\ell}x_{k\ell}^{(\tau)}-\min_{k,\ell}x_{k\ell}^{(\tau)}} .
\label{eq:fntm}
\end{equation}
The minimum and maximum are computed over the 144 OD entries of the same frame.
Unlike the image-oriented scaling in~\cite{sun2026lead}, we keep FN-TM values in $[0,1]$ because our recurrent models operate on vector-valued TM sequences rather than images. 
Due to the inversion in~\cref{eq:fntm}, larger FN-TM values indicate
lower raw traffic relative to the same frame. All losses and reported errors are therefore computed in FN-TM space.

We construct chronological sliding windows for direct multi-step forecasting. Each input $\mathbf{X}_{i}\in\mathbb{R}^{24\times144}$ contains 24 frames, or two hours of history, and each target
$\mathbf{Y}_{i}\in\mathbb{R}^{20\times144}$ contains the next 20 frames, or 100 minutes. 
A model predicts all 20 horizons in one forward pass, so RMSE@1,
RMSE@10, and RMSE@20 are slices of the same direct forecast tensor rather than auto-regressive roll-outs; they correspond to 5, 50, and 100 minutes ahead, respectively. 
Windows are split chronologically into training, validation, and test sets with a 70/15/15 ratio.

All models are trained for 50 epochs with mean squared error (MSE) loss, learning rate
$10^{-3}$, and five random seeds with Adam optimizer~\cite{kingma2014adam}. 
The shared learning rate and epoch budget are
intentional: we compare fixed-budget recurrent models rather than independently tuning for each architecture. 
For each seed $s$, pooled RMSE is
$\mathrm{RMSE}_{\mathrm{pooled}}^{(s)}
=\sqrt{(N_{\mathrm{test}}HC)^{-1}\sum_{i,h,c}
(\hat{y}_{i,h,c}^{(s)}-y_{i,h,c})^2}$, where $H=20$ and $C=144$. Thus, squared errors are pooled over all test windows, direct horizons, and OD channels before a single square root is taken. 
The table reports mean$\pm$std over the five seed-level pooled RMSE values. Horizon-specific RMSE is computed analogously by fixing the horizon and pooling over test windows and OD channels.

We also report OD-channel wins to summarize localized behavior. For each OD channel, we compute the seed-averaged per-channel RMSE after pooling over test windows and horizons, and assign the channel to the model with the lowest value.
To summarize validation convergence under the fixed training budget, we report the normalized Val-loss AULC~\cite{viering2022shape,mazzoni2004active}. 
For model $m$, seed $s$, validation loss $\ell_{m,s,e}$ at epoch $e$, and $E=50$ epochs, we compute
\begin{equation}
\mathrm{AULC}_{\mathrm{val}}^{(s)}(m)=
\frac{1}{E-1}
\sum_{e=1}^{E-1}
\frac{\ell_{m,s,e}+\ell_{m,s,e+1}}{2}.
\label{eq:valaulc}
\end{equation}
We report mean$\pm$std over the five seed-level values. Since epochs are equally
spaced, \cref{eq:valaulc} is the trapezoidal-rule average validation loss across
training. Lower values indicate that a model maintained lower validation loss
within the fixed budget; this metric is used as a convergence summary, not as a
test-set accuracy metric.

\section{Results and Analysis}

\begin{table*}[t]
\centering
\caption{Parameter count and RMSE for direct Abilene whole-matrix forecasting in
FN-TM space. Results are mean$\pm$std over five seeds. Bold and underlining
indicate the lowest and second-lowest errors in each RMSE column, respectively.}
\label{tab:main-results}
\scriptsize
\resizebox{\textwidth}{!}{%
\begin{tabular}{@{}lrrrrr@{}}
\toprule
Model & Params & RMSE@1 & RMSE@10 & RMSE@20 & RMSE pooled \\
\midrule
LSTM-L & 36,624 & \underline{0.06414$\pm$0.00061} & \textbf{0.06961$\pm$0.00007} & \underline{0.07158$\pm$0.00023} & \underline{0.06920$\pm$0.00019} \\
LSTM-S & 8,904 & 0.06730$\pm$0.00066 & 0.07201$\pm$0.00052 & 0.07326$\pm$0.00049 & 0.07155$\pm$0.00053 \\
G-FWP & 7,256 & 0.06768$\pm$0.00068 & 0.07043$\pm$0.00055 & 0.07233$\pm$0.00049 & 0.07038$\pm$0.00057 \\
\midrule
G-QKANFWP & 8,189 & \textbf{0.06229$\pm$0.00053} & \underline{0.06979$\pm$0.00031} & \textbf{0.07130$\pm$0.00025} & \textbf{0.06897$\pm$0.00030} \\
GQKAN-FWP & 7,145 & 0.06533$\pm$0.00046 & 0.07165$\pm$0.00024 & 0.07249$\pm$0.00022 & 0.07082$\pm$0.00024 \\
GQKAN-QKANFWP & 4,637 & 0.06593$\pm$0.00052 & 0.07194$\pm$0.00031 & 0.07284$\pm$0.00037 & 0.07117$\pm$0.00035 \\
\bottomrule
\end{tabular}
}
\end{table*}

\subsection{Aggregate Accuracy and Parameter Efficiency}

\Cref{tab:main-results} shows that all three quantum-inspired variants improve
over the matched-size LSTM-S in pooled RMSE. The strongest result is obtained by
G-QKANFWP, which reaches $0.06897\pm0.00030$ pooled RMSE with 8,189 parameters.
This is lower than LSTM-S ($0.07155\pm0.00053$) despite using slightly fewer
parameters, and it is also slightly lower than the larger LSTM-L
($0.06920\pm0.00019$) while using only 22.4\% of LSTM-L's parameter count. The
absolute margin over LSTM-L is small, but the size difference is substantial,
corresponding to a 77.6\% parameter reduction.

The comparison with G-FWP is central to the interpretation. G-FWP retains the
gated fast-weight framework but removes the HQKAN components, and its pooled
RMSE is $0.07038\pm0.00057$. G-QKANFWP therefore does not win merely because it
uses a gated fast-weight structure; the quantum-inspired fast readout provides a
measurable gain under the same recurrent modeling family. The other two
quantum-inspired variants are also better than LSTM-S in pooled RMSE, but they
trail G-QKANFWP and LSTM-L. This suggests that HQKAN placement matters: on this
TM task, a classical slow programmer combined with a HQKAN fast readout is more
effective than placing HQKAN only in the slow programmer or on both sides.

Horizon-level results further refine the conclusion. G-QKANFWP is best at
$H=1$ and $H=20$, whereas LSTM-L remains slightly better at $H=10$. At $H=1$,
G-QKANFWP improves over LSTM-S by about 7.4\% relative and over LSTM-L by about
2.9\% relative. At $H=20$, the advantage over LSTM-L is much smaller but still
favors G-QKANFWP. Therefore, the correct claim is not universal dominance over a
larger LSTM, but a favorable accuracy--parameter trade-off under a fixed
training budget.

\begin{figure*}[t]
\centering
\includegraphics[width=0.95\textwidth]{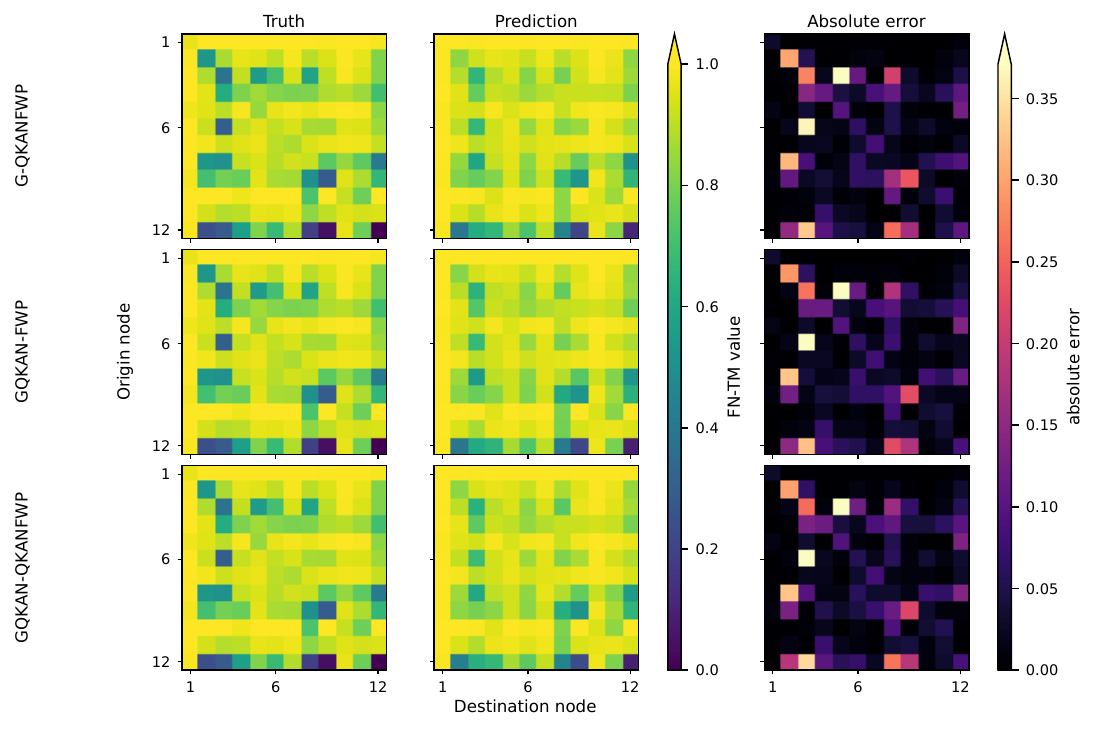}
\vspace{-15pt}
\caption{A qualitative $t+20$ matrix example for test window 275 in normalized
FN-TM space. Panels compare the ground truth, model prediction, and absolute
error.}
\label{fig:matrix-example}
\vspace{-15pt}
\end{figure*}

\Cref{fig:matrix-example} provides a qualitative $H=20$ matrix diagnostic in
FN-TM space. The heatmaps are useful for checking whether low scalar error
corresponds to coherent matrix-level structure rather than only average-error
reduction. We treat this example as supporting evidence; the primary evidence
remains the five-seed aggregate results in \cref{tab:main-results}.

\begin{table}[t]
\centering
\caption{Performance comparison of OD-channel wins and Val-loss AULC across evaluated models. An OD channel win is attributed to the model achieving the lowest pooled per-channel RMSE. Lower Val-loss AULC indicates faster and more sustained loss reduction during training.}
\vspace{-5pt}
\label{tab:channel_auc}
\scriptsize
\resizebox{\columnwidth}{!}{%
\begin{tabular}{@{}lrrr@{}}
\toprule
Model & OD wins & OD win rate & Val-loss AULC \\
\midrule
LSTM-L & 52 / 144 & 36.1\% & 0.00299$\pm$0.00001 \\
LSTM-S & 8 / 144 & 5.6\% & 0.00332$\pm$0.00002 \\
G-FWP & 8 / 144 & 5.6\% & 0.00353$\pm$0.00010 \\
\midrule
G-QKANFWP & 33 / 144 & 22.9\% & 0.00298$\pm$0.00001 \\
GQKAN-FWP & 39 / 144 & 27.1\% & 0.00322$\pm$0.00002 \\
GQKAN-QKANFWP & 4 / 144 & 2.8\% & 0.00326$\pm$0.00001 \\
\bottomrule
\end{tabular}
}
\vspace{-15pt}
\end{table}

\subsection{Validation Convergence and OD-Channel Wins}

\Cref{tab:channel_auc} shows that all three quantum-inspired variants have lower
val-loss AULC than LSTM-S and G-FWP. This indicates more favorable
fixed-budget training curves for the quantum-inspired family, not only lower
final test RMSE. G-QKANFWP has the lowest Val-loss AULC,
$0.00298\pm0.00001$, and is essentially tied with LSTM-L
($0.00299\pm0.00001$). Thus, the convergence result is consistent with the
aggregate RMSE result: G-QKANFWP is the most effective compact recurrent model
in this comparison, while LSTM-L remains a strong large recurrent baseline.

The OD-channel win counts reveal a complementary pattern. LSTM-L wins the most
channels overall, 52 of 144, but G-QKANFWP and GQKAN-FWP win 33 and 39 channels,
respectively, far more than LSTM-S and G-FWP, which each win only 8. This means
that the quantum-inspired models are not only reducing pooled error through a
small set of high-variance channels; they also provide localized advantages over
the matched recurrent baselines. At the same time, GQKAN-FWP wins more channels
than G-QKANFWP despite having weaker pooled RMSE, suggesting that GQKAN-FWP may
help on a broader set of lower-impact OD flows, whereas G-QKANFWP better reduces
the aggregate error that dominates the pooled metric.

\begin{figure}[t]
\centering
\includegraphics[width=0.95\columnwidth]{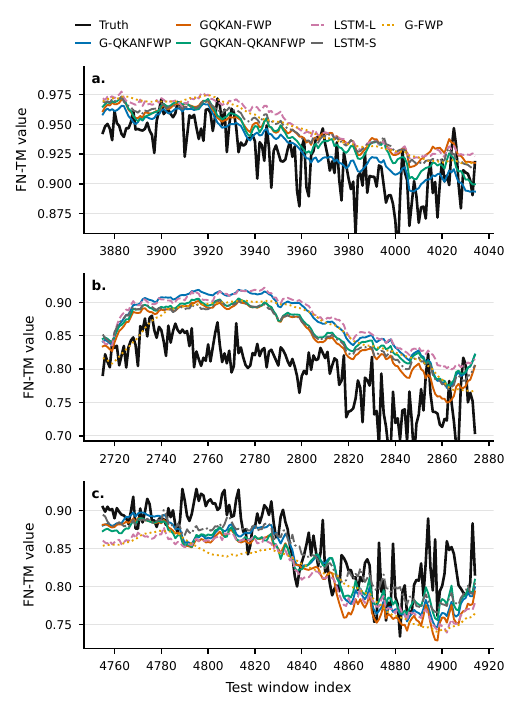}
\vspace{-15pt}
\caption{Prediction traces at t+20 for selected OD channels: (a) G-QKANFWP, channel 14; (b) GQKAN-FWP, channel 142; and (c) GQKAN-QKANFWP, channel 92.}
\label{fig:prediction-traces}
\vspace{-15pt}
\end{figure}

\Cref{fig:prediction-traces} gives fixed-horizon traces for selected OD channels
where a quantum-inspired model improves over its strongest classical comparator.
Each point is one sliding test window evaluated at $H=20$. These traces provide
an interpretability diagnostic for the localized channel-win results and should
be read alongside the aggregate metrics rather than as standalone evidence.

The relative ordering also differs from the broader pattern observed in the
prior gated QKAN-FWP study, where GQKAN-QKANFWP was often the most stable across
heterogeneous tasks \cite{peng2026gatedqkanfwp}. Abilene whole-matrix
forecasting is different: the model must predict many spatially coupled OD
channels, and we intentionally keep a common learning rate for all models to
emphasize fixed-budget training. Under this protocol, the fully HQKAN-based and
most compressed variant remains competitive with LSTM-S but does not dominate.
This motivates a more detailed future study of learning-rate sensitivity,
training dynamics, and the interaction between HQKAN placement and spatial
cross-channel structure.

Overall, the evidence supports a focused conclusion. For parameter-efficient
recurrent TM forecasting, G-QKANFWP is the best candidate among the evaluated
models: it beats the matched-size LSTM-S, outperforms the classical G-FWP
baseline, slightly leads the larger LSTM-L in pooled RMSE, and matches LSTM-L in
Val-loss AULC while using far fewer parameters. The broader
quantum-inspired family also shows strong fixed-budget convergence and
meaningful channel-level gains, although LSTM-L remains the strongest model by
OD-channel win count.

\section{Conclusion}

We studied quantum-inspired fast-weight programmers for direct Abilene
traffic-matrix forecasting under a resource-conscious recurrent comparison. The
main result is that G-QKANFWP achieves the best pooled RMSE among the evaluated
models while using only 22.4\% of the parameters of the larger LSTM-L. All three quantum-inspired variants outperform the matched-size
LSTM-S in pooled RMSE and show lower Val-loss AULC than both LSTM-S and
the classical G-FWP baseline. The G-FWP comparison is especially important: it
shows that the result is not solely due to the gated fast-weight framework, but
also to the quantum-inspired module placement. On this task, placing HQKAN in the
fast readout with a classical slow programmer gives the best aggregate accuracy.

The evidence is strongest when interpreted as an accuracy--efficiency result.
LSTM-L remains competitive: it is slightly better at $H=10$ and wins the largest
number of OD channels. However, G-QKANFWP leads LSTM-L in pooled RMSE, matches it in Val-loss AULC, and requires far fewer parameters. These findings suggest that G-QKANFWP is a strong compact recurrent candidate for TM forecasting when model size and training budget matter.

This study intentionally isolates the recurrent temporal model rather than
combining it with a dedicated spatial module. A natural next step is to pair
G-QKANFWP with graph, diffusion, or topology-aware components that explicitly
model OD-channel dependencies while leaving the recurrent fast-weight module to
capture temporal evolution. Further work should also conduct a detailed
hyperparameter study, especially learning-rate sensitivity across the gated
QKAN-FWP family, and analyze training dynamics in greater depth. Finally,
production-oriented evaluation should extend beyond FN-TM space to
online-compatible normalization, raw-scale forecasting, additional network
benchmarks, and measured inference cost, memory movement, and energy use on
edge or network-control platforms.

\section*{Acknowledgment}
K.-C. Peng, J.-C. Jiang, and C.-H. Lin thank the National Center for High-Performance Computing (NCHC), National Institutes of Applied Research (NIAR), Taiwan, for providing computational and storage resources supported by the National Science and Technology Council (NSTC), Taiwan, under Grants No. NSTC 114-2119-M-007-013.

\clearpage
\bibliographystyle{IEEEtran}
\bibliography{references}

\end{document}